\documentclass[twocolumn,english,showpacs,preprintnumbers,amsmath,amssymb,floatfix]{revtex4}
\usepackage[T1]{fontenc}
\usepackage[latin9]{inputenc}
\usepackage{color}
\usepackage{array}
\usepackage{amstext}
\usepackage{graphicx}
\usepackage{esint}

\makeatletter

\@ifundefined{textcolor}{}
{%
 \definecolor{BLACK}{gray}{0}
 \definecolor{WHITE}{gray}{1}
 \definecolor{RED}{rgb}{1,0,0}
 \definecolor{GREEN}{rgb}{0,1,0}
 \definecolor{BLUE}{rgb}{0,0,1}
 \definecolor{CYAN}{cmyk}{1,0,0,0}
 \definecolor{MAGENTA}{cmyk}{0,1,0,0}
 \definecolor{YELLOW}{cmyk}{0,0,1,0}
 }

\@ifundefined{definecolor}
 {\usepackage{color}}{}
\@ifundefined{definecolor}
 {\usepackage{color}}{}
\makeatother

\makeatother

\usepackage{babel}

\begin{document}

\title{Proton fragmentation functions considering finite-mass corrections}

\author{S. M. Moosavi Nejad$^{a,b}$}
\email{mmoosavi@yazd.ac.ir}

\author{M. Soleymaninia$^{c}$}
\email{maryam_soleymaninia@ipm.ir}

\author{A. Maktoubian$^{a}$}

\affiliation{ 
$^{(a)}$Faculty of Physics, Yazd University, P.O.Box 89195-741, Yazd, Iran\\
$^{(b)}$ School of Particles and Accelerators, Institute for Research
in Fundamental Sciences (IPM), P.O.Box 19395-5531, Tehran, Iran\\
$^{(c)}$ Department of Physics, Payame Noor Universtiy, P.O.Box 19395-3697, Tehran, Iran
}
\date{\today}

\begin{abstract}

We present new sets of proton fragmentation functions (FFs) 
describing the production of protons from the gluon and each of the quarks, obtained by 
a global fit to all relevant data sets of
single-inclusive electron-positron annihilation.
Specifically, we determine their uncertainties using the Gaussian method for error estimation.
Our analysis is in good agreement with the $e^+e^-$ annihilation data.
We also include finite-mass effects of the proton in our calculations, 
a topic with  very little attention paid to in the literatures. 
Proton mass effects turn out to be appreciable
for gluon and light quark FFs. The
inclusion of finite-mass effects tends to improve the overall description of the data
by reducing the minimized $\chi^2$ values significantly.
As an application, we apply the extracted FFs to make predictions for
the scaled-energy distribution of protons 
 inclusively produced in  top quark decays at next-to-leading order,
relying on the universality and scaling violations of FFs.
\end{abstract}


\maketitle

\section{Introduction}
\label{sec1}

Fragmentation functions $D_i^H(z, \mu_F^2)$, describe the probability for
a parton $\it{i}$ at the factorization scale $\mu_F$ to fragment into a hadron $\it{H}$
carrying away a fraction $\it{z}$ of its momentum. They are 
key quantities  for calculating hadroproduction cross sections. 
In this respect, one needs to determine these functions with high accuracy.
Recently, in \cite{Anderle:2015lqa} pion fragmentation functions (FFs)
are studied  at next-to-next-to-leading  order (NNLO)  using  data from $e^+e^-$ annihilation.\\
The specific importance of FFs is due to  their model-independent predictions of  cross sections at 
the Large Hadron Collider (LHC).
Specially, to study the properties of top quarks at the
LHC, one of the proposed channels is to consider the energy spectrum of outgoing hadrons {\it H} from top decays in the process $t\rightarrow W^+b(\rightarrow H)$.
For this purpose, having parton-level differential decay rates for the subprocess $t\rightarrow bW(g)$ at NLO \cite{Nejad:2013fba, Kniehl:2012mn, Corcella:2001hz} 
and FFs of partons into hadrons $D_{i}^H$ ({\it i=b,g}), one can calculate the energy distribution of observed hadrons.
For example, the CMS collaboration reconstructs the top mass from the peak of the energy distribution of  bottom-flavored hadrons \cite{CMS:2015jwa,Corcella:2015kth}.
Since the hadronization mechanism is universal and independent of the perturbative processes
which produce the partons one can exploit, for example, the existing data on $e^+e^-\rightarrow b\bar b\rightarrow H+jets$
events to fit the proposed models for the FFs describing $b\rightarrow H$, and 
use them to make predictions for 
other processes, such as top decays $t\rightarrow W^+(\rightarrow l^+\nu_l)+b(\rightarrow H)$.
FFs are also useful for theoretical calculations of inclusive hadron production
at the Relativistic Heavy Ion Collider (RHIC) and for
investigating the origin of the proton spin.

Generally, there are two main approaches to evaluate the FFs. 
The first approach is based on the fact that the FFs for hadrons containing a heavy quark or antiquark can be computed
theoretically using perturbative QCD (pQCD) \cite{Ma:1997yq,Braaten:1993rw}.
The first theoretical attempt to explain the procedure of hadron
production from a heavy quark was made by Bjorken \cite{Bjorken:1977md} by using a naive quark-parton model. He deduced that
the inclusive distribution of heavy hadrons should peak almost at $z = 1$, where $z$ refers to the scaled-energy variable of hadrons.
The pQCD framework was applied by Peterson \cite{Peterson:1982ak}, Suzuki \cite{Suzuki:1977km}, 
Amiri and Ji \cite{Amiri:1986zv}, while in this framework
Suzuki calculates the heavy FFs by using a convenient Feynman diagram and considering
a typical wave function for the hadronic bound states. One of us, using Suzuki's approach, has calculated
the FFs for a charm quark to split into S-wave $D^0/D^+$ mesons \cite{Nejad:2013vsa} and for a gluon to split into S-wave charmonium states ($\eta_c, J/\psi$) \cite{MoosaviNejad:2015dia} at leading order in the QCD coupling constant $\alpha_s$.\\
In the second approach which is frequently used to obtain the
FFs, these functions are parametrized in a specific model and
extracted from experimental data analysis using the data from 
$e^+e^-$  reactions, lepton-hadron and hadron-hadron scattering processes.
This situation is very similar to the determination of the parton distribution functions (PDFs)\cite{Arbabifar:2013tma}.
Among all scattering processes, the
best processes which provide a clean environment to determine the FFs are $e^+e^-$ annihilation
processes. However, without an initial hadron state one can not separate quark distributions
from antiquark distributions.
There are several theoretical studies on QCD analysis
of FFs which used special parameterizations and different experimental data in their global analysis.
Recent extracted FFs for light hadrons are related to AKK \cite{Albino:2008fy}, SKMA \cite{Soleymaninia:2013cxa,Soleymaninia:2014oya}, LSS \cite{Leader:2015hna}, DSEHS \cite{deFlorian:2014xna}, 
DSS \cite{deFlorian:2007aj,deFlorian:2007hc} and HKNS \cite{Hirai:2007cx} collaborations who used different phenomenological models and varying experimental data.
In  \cite{Soleymaninia:2013cxa,Soleymaninia:2014oya}, we determined the $\pi^\pm$ and $K^\pm$ FFs,
both at leading order (LO) and NLO through a global fit 
to the single-inclusive $e^+e^-$ annihilation (SIA) data and the
semi-inclusive deep inelastic scattering  (SIDIS) asymmetry data from
HERMES and COMPASS. 
There, we broke the symmetry assumption between the quark and antiquark FFs for favored partons by using
the asymmetry data.\\
In this paper, we present a new functional form of proton FFs up to NLO, 
obtained through a global fit to SIA data from LEP (ALEPH and DELPHI collaborations),
SLAC (SLD and TPC collaborations) and DESY (TASSO collaboration)  \cite{aleph91,delphi91,delphi91-2,sld91,tpc29,tasso34_44}.
 We also impose the effects of the proton mass on the FFs, 
a topic with very little attention paid to in the literatures, e.g. \cite{Soleymaninia:2013cxa,Soleymaninia:2014oya,Leader:2015hna,deFlorian:2014xna,deFlorian:2007aj,deFlorian:2007hc,Hirai:2007cx}. We  show
that this  effect is important in calculating gluon and light quark FFs, specifically
at low energies. This effect reduces the minimized $\chi^2$ and
makes a more convenient fit.
 
In the Standard Model (SM) of particle physics the top quark has a short
lifetime ($\tau_t\approx 0.5\times 10^{-24} s$ \cite{Chetyrkin:1999ju}), so it decays before hadronization takes place. Therefore its full polarization content is retained
when it decays. As was mentioned, one way to study its properties is to consider the
energy distributions of the produced hadrons in top decays. Here, we also make predictions for the
scaled-energy distribution of protons inclusively produced in 
top quark decays, $t\rightarrow W^++b(\rightarrow p+X)$, using the extracted FFs at our analysis 
and the parton-level Wilson coefficients calculated in  \cite{Kniehl:2012mn, Corcella:2001hz}. 
These predictions will also enable us to deepen our understanding of the nonperturbative aspects
of proton formation by hadronization and to pin down the proton FFs.

This paper is organized as follows. In section~\ref{sec2} we describe the formalism and
our parametrization of proton fragmentation functions at LO and NLO. Using the SIA data we present our fit results
and shall compare our results against FFs from other well-known collaborations.
We also describe the Gaussian method to calculate errors in our analysis.
 In section~\ref{sec3}, we explain 
how to impose the effect of the hadron mass on the FFs and compare our results in both cases.
 Our predictions for the energy spectrum of protons
produced in unpolarized top quark decays are presented in section~\ref{sec4}.
Our results are summarized in section~\ref{sec5}.

\section{QCD analysis of proton fragmentation functions}
\label{sec2}
\subsection{Formalism}
\begin{table*}[th]
	\caption{\label{tab:exLO}The individual $\chi^2$ values at LO for
		each collaboration and the total $\chi^2$ fit for proton.}
	\begin{ruledtabular}
		\tabcolsep=0.06cm \footnotesize
		\begin{tabular}{lccccc}
			collaboration& data & $\sqrt{s}$  GeV&  data   &  $\chi^2$(LO)\\
			& properties &          &  points & normalization in fit&  \\\hline
			TPC \cite{tpc29}  &   untagged   & 29      & 8  & 10.959  \\
			TASSO \cite{tasso34_44}  & untagged \  &  34 & 4 & 2.006\\
			ALEPH \cite{aleph91}    & untagged  & 91.2 & 18 &  18.211 \\
			SLD \cite{sld91}  & untagged\  & 91.28 & 28  & 80.643\\
			& $uds$ tagged         &  91.28 & 29 & 61.263 \\
			& $c$ tagged           &  91.28 & 29 & 40.286\\
			& $b$ tagged           &  91.28 & 28 & 60.358\\
			DELPHI \cite{delphi91,delphi91-2}  & untagged & 91.2  & 17 & 2.990 \\
			& $uds$ tagged         &  91.2 & 17 & 4.710 \\
			& $b$ tagged           &  91.2 & 17 & 14.681 \\
			{\bf TOTAL:} & & &195 &296.107\\
			($\chi^{2}$/ d.o.f ) & & & &1.711\\
		\end{tabular}
	\end{ruledtabular}
\end{table*}
\begin{table*}[th]
	\caption{\label{tab:exNLO} As in Table \ref{tab:exLO}, but at NLO.}
	\begin{ruledtabular}
		\tabcolsep=0.06cm \footnotesize
		\begin{tabular}{lccccc}
			collaboration& data & $\sqrt{s}$  GeV&  data   &  $\chi^2$(NLO)\\
			& properties &          &  points & normalization in fit &   
			\\\hline
			TPC \cite{tpc29}  &   untagged   & 29      & 8  & 21.984  \\
			TASSO \cite{tasso34_44}  & untagged \  &  34 & 4 & 2.143\\
			ALEPH \cite{aleph91}    & untagged  & 91.2 & 18 &  19.955 \\
			SLD \cite{sld91}  & untagged\  & 91.28 & 28  &83.960\\
			& $uds$ tagged         &  91.28 & 29 & 43.920 \\
			& $c$ tagged           &  91.28 & 29 & 39.010\\
			& $b$ tagged           &  91.28 & 28 & 45.795\\
			DELPHI \cite{delphi91,delphi91-2}  & untagged & 91.2  & 17 & 4.734 \\
			& $uds$ tagged         &  91.2 & 17 & 5.861 \\
			& $b$ tagged           &  91.2 & 17 & 13.929 \\
			{\bf TOTAL:} & & &195 &281.291\\
			($\chi^{2}$/ d.o.f ) & & & &1.626\\
		\end{tabular}
	\end{ruledtabular}
\end{table*}
\begin{table*}[th]
	\caption{\label{tab:loproton} Values of the fit parameters for the  proton FFs at LO
		in the starting scales $\mu_0$.}
	\centering
	\hspace{1cm}
	\begin{tabular}{cccccc}
		\hline
		\hline
		flavor $i$ &$N_i$ &\quad $\alpha_i$ & $\beta_i$ &$\gamma_i$\\
		\hline
		$u$ &$0.028\pm 3.105$&$-1.547\pm 1.218$&$2.710\pm 0.657$&$0.052\pm 0.014$ 
		\\
		$d$ &$3.382\pm 1.185$&$-1.547\pm 1.218$&$2.710\pm 0.657$&$0.052\pm 0.014$ 
		\\
		$ \overline{u}$&$7.634\pm 1.481$&$-0.053\pm 0.021$&$5.389\pm 0.614$&$1.269\pm 0.618$  
		\\
		$ \overline{d}$&$8.592\pm 3.268$&$-0.053\pm 0.021$&$5.389\pm 0.614$&$1.269\pm 0.618$  
		\\
		$ s, \overline{s}$&$3.110\pm 2.009$&$-0.053\pm 0.021$&$5.389\pm 0.614$&$1.269\pm 0.618$  
		\\
		$c, \overline{c}$&$4.829\pm 2.700$&$0.733\pm 0.556$&$7.046\pm 2.093$&$...$
		\\
		$b, \overline{b}$&$10.031\pm 3.430$&$0.681\pm 0.282$&$11.466\pm 1.360$&$...$ 
		\\
		$g$&$1.237\pm 0.331$&$6.915\pm 2.380$&$0.482\pm 0.191$&$...$             
		\\
		\hline
		\hline
	\end{tabular}
\end{table*}
\begin{table*}[th]
	\caption{\label{tab:nloproton} As in Table \ref{tab:loproton}, but at NLO.}
	\centering
	\hspace{1cm}
	\begin{tabular}{cccccc}
		\hline
		\hline
		flavor $i$ &$N_i$ & $\alpha_i$ & $\beta_i$ &$\gamma_i$\\
		\hline
		$u$ &$3.206\pm 2.852$&$-0.759\pm 0.179$&$17.935\pm 2.859$&$2.320\pm 0.225$ 
		\\
		$d$ &$3.284\pm 1.686$&$-0.759\pm 0.179$&$17.935\pm 2.859$&$2.320\pm 0.225$ 
		\\
		$ \overline{u}$&$21.005\pm 2.071$&$-0.533\pm 0.127$&$3.795\pm 1.281$&$0.1007\pm 0.035 $  
		\\
		$ \overline{d}$&$15.990\pm 2.352$&$-0.533\pm 0.127$&$3.795\pm 1.281$&$0.1007\pm 0.035$  
		\\
		$ s, \overline{s}$&$7.999\pm 1.580$&$-0.533\pm 0.127$&$3.795\pm 1.281$&$0.1007\pm 0.035$
		\\
		$c, \overline{c}$&$6.120\pm 1.575$&$0.826\pm 0.217$&$7.949\pm 1.957$&$...$
		\\
		$b, \overline{b}$&$7.218\pm 1.605$&$0.560\pm 0.108$&$11.617\pm 2.307$&$...$ 
		\\
		$g$&$3.739\pm 1.862$&$4.290\pm 0.814$&$1.736\pm 0.218$&$...$   
		\\
		\hline
		\hline
	\end{tabular}
\end{table*}
FFs are nonperturbative functions describing the hadronization processes, so they
have an important role in the calculation of single-inclusive hadron production in any reaction.
According to the factorization theorem of the parton model \cite{Collins:1998rz}, 
the leading twist component of  inclusive production measurement of any single hadron can be written as the convolution of perturbative partonic cross sections, with nonperturbative FFs and PDFs  
to account for any hadrons in the initial and final states, respectively. As an example, the cross section for the production of hadron $H$ in the typical 
scattering process $A+B\rightarrow H+X$, can be expressed as \cite{Collins:1987pm}
\begin{eqnarray}
d\sigma&=&\sum_{a,b,c}\int_0^1 dx_a\int_0^1 dx_b\int_0^1 dz f_{a/A}(x_a, Q^2)f_{b/B}(x_b, Q^2)\nonumber\\
&&\times d\hat\sigma(a+b\rightarrow c+X)D_c^H(z, Q^2),
\end{eqnarray}
where  $\it{a}$ and $\it{b}$ are incident partons in the colliding initial hadrons $A$ and $B$, respectively,
$f_{a/A}$ and $f_{b/B}$ are the PDFs at the scale $Q^2$ of the partonic subprocess $a+b\rightarrow c+X$, 
 $\it{c}$ is the
fragmenting parton (either a gluon or a quark) and $X$ stands for the unobserved jets. 
Here, $D_c^H(z, Q^2)$ is the FF at the scale $Q^2$ which
can be obtained by evolving from the initial FF $D_c^H(z, Q_0^2)$
using the Dokshitzer-Gribov-Lipatov-Altarelli-Parisi (DGLAP) renormalization group equations \cite{dglap}.\\
Since we  determine the FFs from single-inclusive hadron production data
in $e^+e^-$ annihilation, we do not need
to deal with uncertainties induced by PDFs as in hadron collisions.
The cross section for the single-inclusive $e^+e^-$ annihilation processes, $e^+e^-\rightarrow (\gamma, Z)\rightarrow H+X$, is expressed as follows \cite{Collins:1998rz}
\begin{eqnarray}\label{fac}
	\frac{1}{\sigma_{tot}}\frac{d}{dx_H}\sigma(e^+e^-\rightarrow HX)&=&\nonumber\\
	&&\hspace{-4cm}\sum_i \int_{x_H}^1\frac{dx_i}{x_i} D_i^H(\frac{x_H}{x_i}, \mu_F^2)\frac{1}{\sigma_{tot}}\frac{d\hat\sigma_i}{dx_i}(x_i, \mu_R^2, \mu_F^2),
\end{eqnarray}
where, $\sigma_{tot}$  is the total partonic
cross section at NLO \cite{Kneesch:2007ey} and $D_i^H$ indicates the probability to find the
hadron $H$ generated from a parton $i(=g, u/\bar u, d/\bar d, \cdots)$ with the scaled-energy fraction $x_H=2E_H/\sqrt{s}$ so that  $s=q^2$ stands for the squared of the four-momenta of the intermediate gauge bosons $\gamma$ and $Z$.
In (\ref{fac}), $x_i$ is also defined in analogy to $x_H$
as $x_i=2E_i/\sqrt{s}$ and the $d\hat\sigma_i/dx_i$ are the differential cross sections at the parton level for the $i\bar{i}$ pair-creation subprocesses;  $e^+e^-\rightarrow (\gamma, Z)\rightarrow i\bar i+(g)$,
which can be calculated in perturbative QCD \cite{Kretzer:2000yf,Kniehl:2005de,Binnewies:1994ju}. In the equation above,
$\mu_R$ and $\mu_F$ stand for the renormalization and factorization scales respectively, so one can use two different values for these scales; however, a choice often made consists of setting $\mu_F^2=\mu_R^2=Q^2$ and we shall adopt this convention in our work.\\
 In section~\ref{sec3}, we shall review   the factorization theorem 
in detail and extend it  in the presence of  the hadron mass.\\
There are several different strategies to extract the FFs from data analysis. In the present analysis we adapt
the zero-mass variable-flavor-number (ZM-VFN) scheme \cite{Kneesch:2007ey}.
This scheme works best for high energy scales, where the 
heavy quark masses are set to zero from the start and the nonzero values of the {\it c}- and {\it b}-quark
masses only enter through the initial conditions of the FFs.
In this scheme, the number of active flavors  also increases with 
the flavor thresholds.
\subsection{Parametrization of proton FFs}
In the phenomenological approach, the FFs are parametrized in a convenient functional form at an initial scale $\mu_0$
at each order, i.e. LO and NLO, and then evolved to higher scales using the DGLAP evolution equations to fit to the existing experimental data. Various phenomenological models like Peterson
model \cite{Peterson:1982ak}, Power model \cite{Kniehl:2008zza},
Cascade model \cite{Webber:1983if} etc., have been developed to describe the FFs.
At the initial scale $\mu_0$, we apply the following flexible functional form
for the proton FFs
\begin{equation}
\label{ff-input}
D_i^p(z,\mu_{0}^{2}) =
N_i z^{\alpha_i}(1-z)^{\beta_i} [1-\delta_i e^{-\gamma_i z}],
\end{equation}
which  is an appropriate form for the light hadrons. Here, $z=x_p/x_i=E_p/E_i$ (with $i=g, u, \bar u, d, \bar d, \cdots$) 
is the energy fraction of the parton $i$
which is taken away by the produced proton.
A simple polynomial parametrization with just three parameters controls the small- and large-{\it z} regions so that in (\ref{ff-input}), the power term in {\it z} restricts  the small-{\it z} region and the power term in  (1-{\it z}) emphasizes the large-{\it z} region. To control the medium $z$-region and to improve the accuracy of the global fit the extra term $[1-\delta_i e^{-\gamma_i z}]$ is considered \cite{Soleymaninia:2013cxa}. In order to get the best fit we assume $\delta_i=1$ for light quarks and $\delta_i=0$ for the gluon and heavy quark FFs.
The free parameters $N_i$, $\alpha_i$, $\beta_i$ and $\gamma_i$  in the proposed form 
are determined by minimizing $\chi^2$ for differential cross section and experimental data $(1/\sigma_{tot}\cdot d\sigma^p/dx_p)_{exp}$.
The $\chi^2$ for $k$ data points is defined as
\begin{equation}
\label{eq:chi2}
\chi^2=\sum_{j=1}^k (\frac{E_j-T_j}{\sigma^{E}_{j}})^2.
\end{equation}
Here,  $T_{j}$ and $E_j$ stand for the theoretical results and experimental values of
 $1/\sigma_{tot}\times d\sigma^p/dx_p$, respectively.  $\sigma^{E}_{j}$
is the error of the corresponding experimental data including the statistical and systematical errors.\\
According to the partonic structure of the proton $p(uud)$ and the general functional form presented in (\ref{ff-input}), we consider the following forms for light-quark FFs
\begin{eqnarray}
\label{minus1}
D_{u}^{p}(z,\mu_{0}^{2})&=&N D_{d}^{p}(z,\mu_{0}^{2})= N_u z^{\alpha_u}(1-z)^{\beta_u}(1-e^{-\gamma_u z}),\nonumber\\
D_{\bar{u}}^{p}(z,\mu_{0}^{2})&=&N^\prime D_{\bar{d}}^{p}(z,\mu_{0}^{2})= N_{\bar{u}} z^{\alpha_{\bar{u}}}
(1-z)^{\beta_{\bar{u}}}(1-e^{-\gamma_{\bar{u}} z}),\nonumber\\
D_{s}^{p}(z,\mu_{0}^{2})&=&D_{\bar{s}}^{p}(z,\mu_{0}^{2})=N^{\prime\prime} D_{\bar{u}}^{p}(z,\mu_{0}^{2})\nonumber\\
&&\hspace{-0.5cm}= N_{s} z^{\alpha_{\bar{u}}}
(1-z)^{\beta_{\bar{u}}}(1-e^{-\gamma_{\bar{u}} z}),
\end{eqnarray}
where the assumption $\delta_i=1$ is used.
For the gluon and heavy quark FFs, considering $\delta_i=0$ we apply a power model in our calculations, as 
applied by  DSS \cite{deFlorian:2007hc} and HKNS \cite{Hirai:2007cx} collaborations, i.e.
\begin{eqnarray}
\label{minus2}
D_{g}^{p}(z,\mu_{0}^{2}) &=& N_g z^{\alpha_g}(1-z)^{\beta_g},\nonumber\\
D_{c}^{p}(z,\mu_{0}^{2}) &=& D_{\bar{c}}^{p}(z,\mu_{0}^{2})=N_c z^{\alpha_c}(1-z)^{\beta_c},\nonumber\\
D_{b}^{p}(z,\mu_{0}^{2}) &=& D_{\bar{b}}^{p}(z,\mu_{0}^{2})=N_b z^{\alpha_b}(1-z)^{\beta_b}.
\end{eqnarray}
The assumption $\delta_i=0$ in (\ref{ff-input}) for the gluon and heavy quark FFs makes a more convenient fit by reducing the minimized $\chi^2$-values.
In the equations above, we also considered the isospin symmetry for sea quarks of proton FFs. 
 Considering Eqs.~(\ref{minus1}) and (\ref{minus2}), there are $20$ parameters which must be determined by fitting.
In the reported parameters by  AKK \cite{Albino:2008fy} and HKNS \cite{Hirai:2007cx} some parameters such as {\it N} and $N^\prime$ (\ref{minus1}) are fixed from the beginning, i.e. a simple parameterization form is applied.\\
In our parametrization, the starting  scale $\mu_{0}$ is different for various partons. The value of $\mu_{0}^{2}=1$~GeV$^2$ 
is chosen for the splitting of a gluon and light-quarks into the proton and
for the $c$- and $b$-quarks it is taken to be $\mu_{0}^{2}=m_{c}^{2}$ and $\mu_{0}^{2}=m_{b}^{2}$, respectively.\\
According to the partonic structure of the proton, the proton FFs can be applied  for the anti-proton as
\begin{eqnarray}
\label{antiproton}
D_{i}^{\bar{p}}(z,\mu_{0}^{2}) &=& D_{\bar{i}}^{p}(z,\mu_{0}^{2}).
\end{eqnarray}
\subsection{Experimental data and our analysis}
In our fits of the proton FFs, we consider 
SIA data from LEP ({\it ALEPH} \cite{aleph91} and {\it DELPHI} \cite{delphi91,delphi91-2} collaborations), 
SLAC ({\it SLD} \cite{sld91} and {\it TPC} \cite{tpc29} collaborations) and DESY ({\it TASSO} \cite{tasso34_44}
collaboration). 
The energy scales of the experimental data range from 29 GeV to 91.28 GeV.
Most of the precision $e^+e^-$ annihilation data, however, come from the LEP and SLAC data at the energy scale of $Q^2=M_Z^2$.
In the data sets reported by the DELPHI
and SLD, authors
distinguished between four cases: fragmentation of $\it{u}$, $\it{d}$ and $\it{s}$ quarks, {\it c} quarks only,
{\it b} quarks only, and all five quark flavors (see Tables \ref{tab:exLO} and \ref{tab:exNLO}).\\
Note that in our global analysis of data sets from the ALEPH experiments we deal with data 
in the form of $1/N\times dN^p/dx_p$, where {\it N} is the number of detected events. This fraction is defined as the ratio of the single-inclusive $e^+e^-$ annihilation cross section ($e^+e^-\rightarrow p+X$) in a certain bin of $x_p$ to the totally inclusive rate, i.e. $1/\sigma_{tot}\times d\sigma^p/dx_p$, where
$x_p$ refers to the energy of proton
scaled to the beam energy $\sqrt{s}$, i.e. $x_p=2 E_p/\sqrt{s}$.\\
In Tables \ref{tab:exLO} and \ref{tab:exNLO}, based on  each collaboration the data characteristics, the $\chi^2$ (\ref{eq:chi2}) and
the $\chi^2$ values per degree of freedom ($\chi^2/d.o.f$) are listed.
The obtained values of $\chi^2 /d.o.f$ for the proton at LO and NLO are $1.711$ and $1.626$ in our global fit, respectively. 
In section \ref{sec3}, we will show that this value is reduced when one includes  the effects of the proton mass in the calculations.\\
Our LO and NLO results for the fit parameters in the starting scales $\mu_0^2$ along with the corresponding uncertainties are listed in Tables \ref{tab:loproton} and \ref{tab:nloproton}. The method of
error calculation shall be described in next subsection.

\begin{figure*}
	\begin{center}
		\includegraphics[width=0.7\linewidth,bb=60 60 660 530]{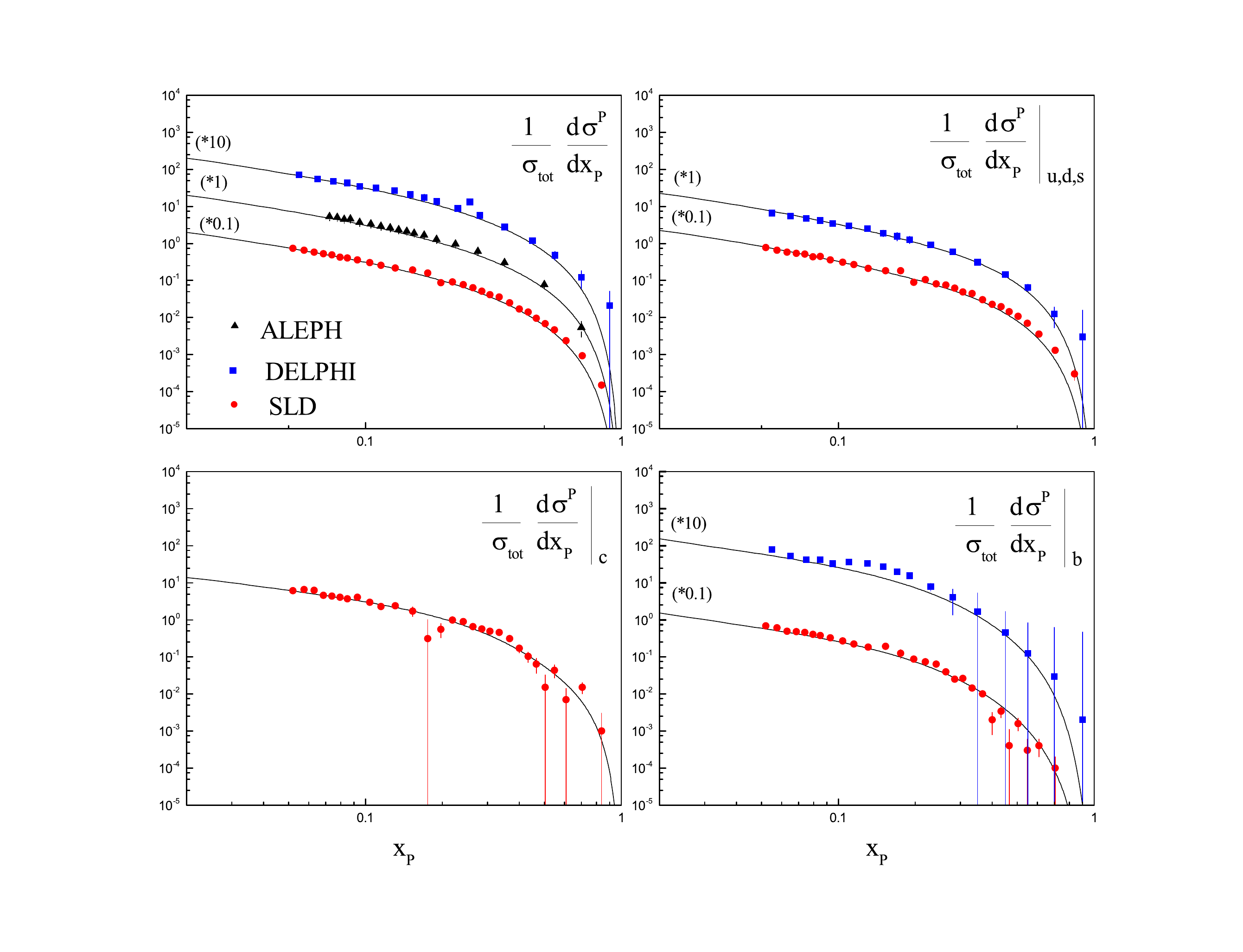}
		\caption{\label{xsec3}%
			Comparison of our NLO results for $1/\sigma_{tot}\times d\sigma/dx_p$ in total and tagged cross sections
			with proton production data at $Q^2=M_Z^2$ by {\it ALEPH} \cite{aleph91}, {\it DELPHI} \cite{delphi91,delphi91-2} and
			{\it SLD} \cite{sld91}.}
	\end{center}
\end{figure*}

\begin{figure*}
	\begin{center}
		\includegraphics[width=0.7\linewidth,bb=60 40 650 530]{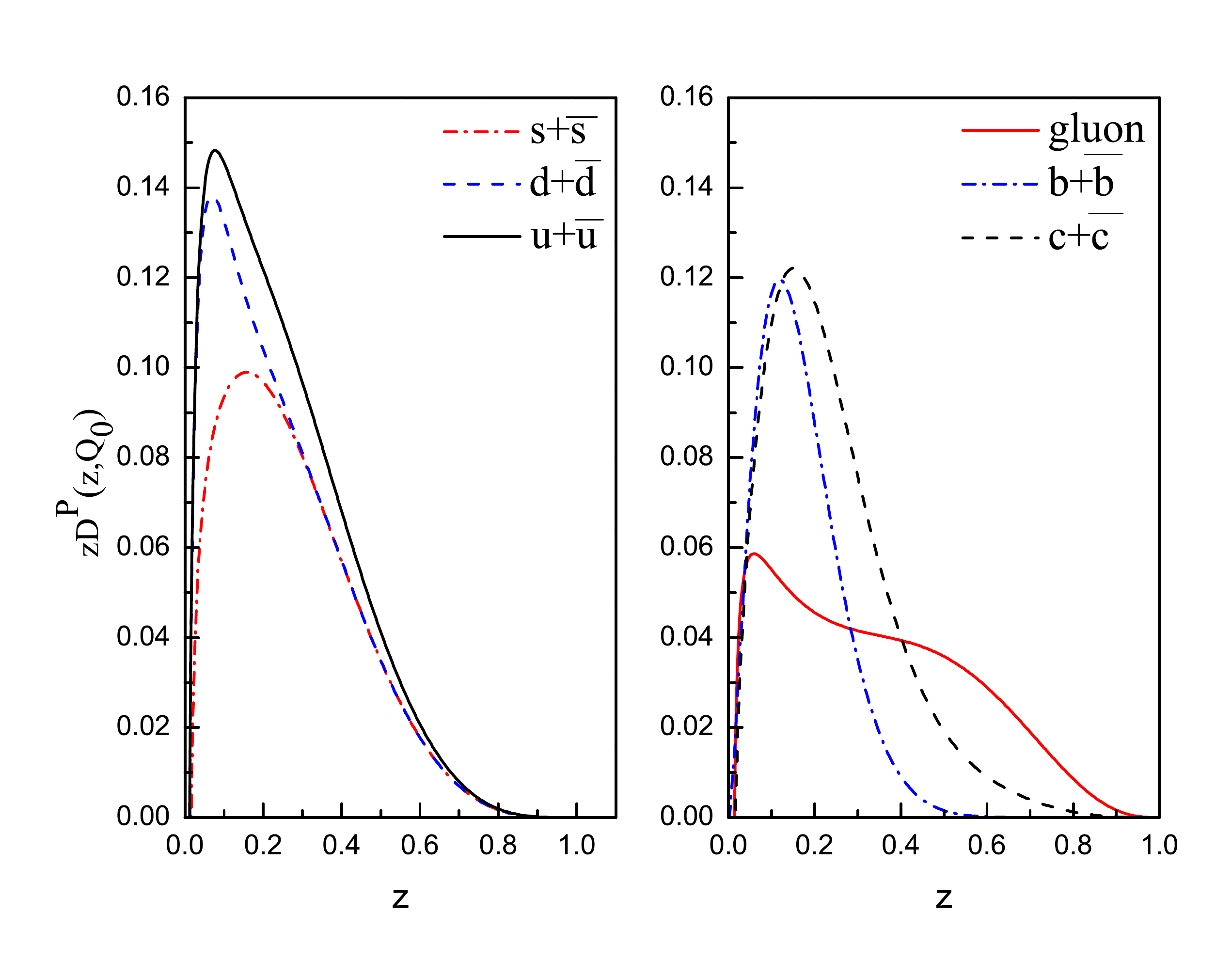}
		\caption{\label{NLOModel3Q0}%
			Proton fragmentation functions  are shown at $Q_0^2=1$ GeV$^2$, $m_c^2$
			and $m_b^2$ at NLO.}
	\end{center}
\end{figure*}

\begin{figure}
\begin{center}
\includegraphics[width=1.2\linewidth,bb=110 70 560 540]{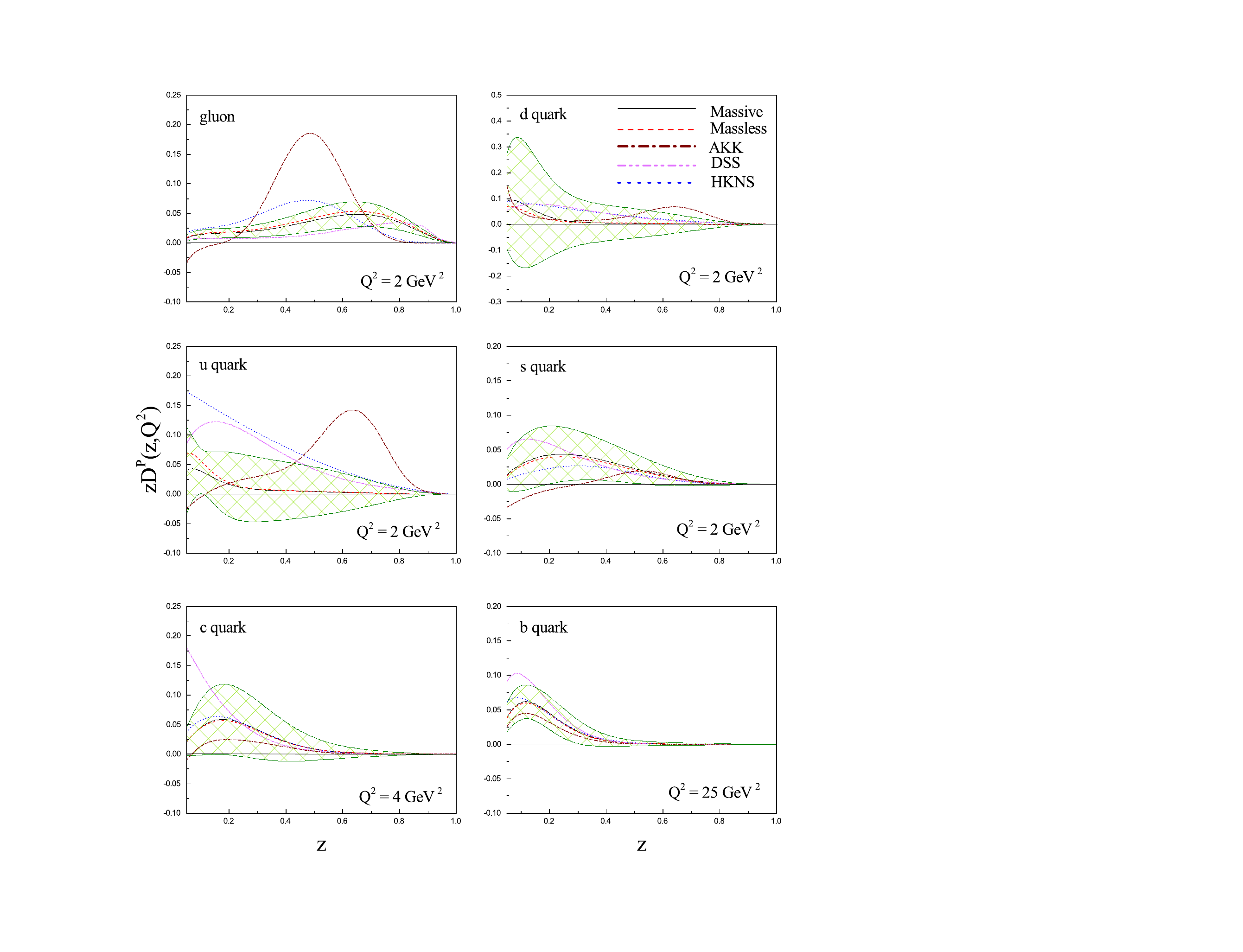}
\caption{\label{plotmodel3}%
The NLO fragmentation densities for the massless (dashed line) 
and massive (solid line) protons are shown at 	
$Q^2=2$ GeV$^2$, $4$ GeV$^2$ and
$25$ GeV$^2$ for the gluon and light quarks, the  {\it c}- and {\it b}-quarks, respectively.
The uncertainties of massive proton FFs are also shown. 
Our results are compared with AKK \cite{Albino:2008fy}, 
DSS  \cite{deFlorian:2007hc} and
HKNS \cite{Hirai:2007cx}  collaborations.}
\end{center}
\end{figure}

In Fig.~\ref{xsec3}, using the NLO results for the proton FFs (Table \ref{tab:nloproton}) and applying the DGLAP equations, we compare
our results for $1/\sigma_{tot}\times d\sigma^p/dx_p$ with SIA experimental data at $\mu^2=M_Z^2$ reported by
the {\it ALEPH} \cite{aleph91}, {\it DELPHI} \cite{delphi91,delphi91-2} and
{\it SLD} \cite{sld91} collaborations.
 In these figures we separate the light, charm and
bottom tagged cross sections, and as it is seen most of the diagrams are in a reliable consistency with the experimental data.
Note that, as Fig.~\ref{xsec3} shows, due to the largeness of the $\chi^2$ contribution for the SLD b-tagged data
(see Table.~\ref{tab:exNLO}) some points are outside of the curves. However, $\chi^2$-values of the heavy
quarks are usually larger than the other data and this might be caused to some extent
by contaminations from weak decay.\\
The NLO proton FFs are presented in Fig.~\ref{NLOModel3Q0} at their corresponding initial scales; 
$ \mu _0^2=1$ GeV$^2$ for the gluon and light quarks, $\mu_0^2=m_c^2$ and $\mu_0^2=m_b^2$ for the {\it c}- and {\it b}-quarks, respectively.
In Fig.~\ref{plotmodel3}, our results for the fragmentation densities
are presented at the scales $Q^2=2$ GeV$^2$ for the gluon and light quarks, and $Q^2=4$ GeV$^2$ and $Q^2=25$ GeV$^2$ for the  {\it c}- and {\it b}-quarks, respectively. Our results for different flavors are compared with the DSS set of proton FFs \cite{deFlorian:2007hc}
that included electron-positron, lepton-nucleon, and hadron-hadron scattering data, and the HKNS FFs \cite{Hirai:2007cx} that included electron-positron data. 
For more comparison we have also applied the AKK extraction of proton FFs
at NLO \cite{Albino:2008fy} that included hadron production data in electron-positron and hadron-hadron scattering data. Comparing our model with other FF models gives a nice all-around description of our model.
As is seen, our results for the heavy quark FFs are in a reliable consistency
with the results obtained by  HKNS collaboration.
In this figure, considering the proton mass effects we also show the massive FFs (solid line) and their error bands at NLO, a topic which is explained in section \ref{sec3}. 
\subsection{The Gaussian method to calculate errors }
\label{errorcalculation}
According to (\ref{minus1}) and (\ref{minus2}), the evolved proton FFs
are  attributive functions of the input parameters which are calculated from the fit,
and their standard linear errors are given by Gaussian error propagation. If $D_{i}^{H}(z,Q^2)$
is the evolved fragmentation density at the scale $Q^2$, then the Gaussian error propagation is defined as
\begin{equation} \label{eq:heserror}
[\delta D_{i}^{H}(z)]^2 =\Delta \chi^2{\sum_{j,k}^{n}\frac{\partial D_{i}^{H}(z,a_j)}{\partial a_j}(H_{jk})^{-1}\frac{\partial  D_{i}^{H}(z,a_k)}{\partial a_k}},
\end{equation}
where $\Delta \chi^2$ is the allowed variation in $\chi^2$, $a_j\mid^{n}_{j=1}$ are fit parameters and $n$ is the number of parameters in the global fit. 
$H_{jk}$ are the elements of the Hessian or covariance matrix of the parameters determined in the QCD analysis at the initial scale $Q^{2}_{0}$ which are defined as
\begin{equation} \label{eq:hessianmatrix}
H_{jk} = \left.\frac{1}{2}\frac{\partial^2\,\chi^2}{\partial a_j\partial a_k}\right|_{\rm min}.
\end{equation}
Consequently, we can calculate the uncertainties of any FFs by using the Hessian  matrix based on the Gaussian method
at any value of $Q^{2}$ by the QCD evolution. 
The results indicate that the FFs, especially gluon and light-quark FFs, have large uncertainties at small $Q^2$.
More information and a detailed discussion can be found in \cite{Soleymaninia:2014oya,Hirai:2007cx}.
\section{Proton mass effects}
\label{sec3}
\begin{figure}
	\begin{center}
		\includegraphics[width=1\linewidth,bb=80 65 500 540]{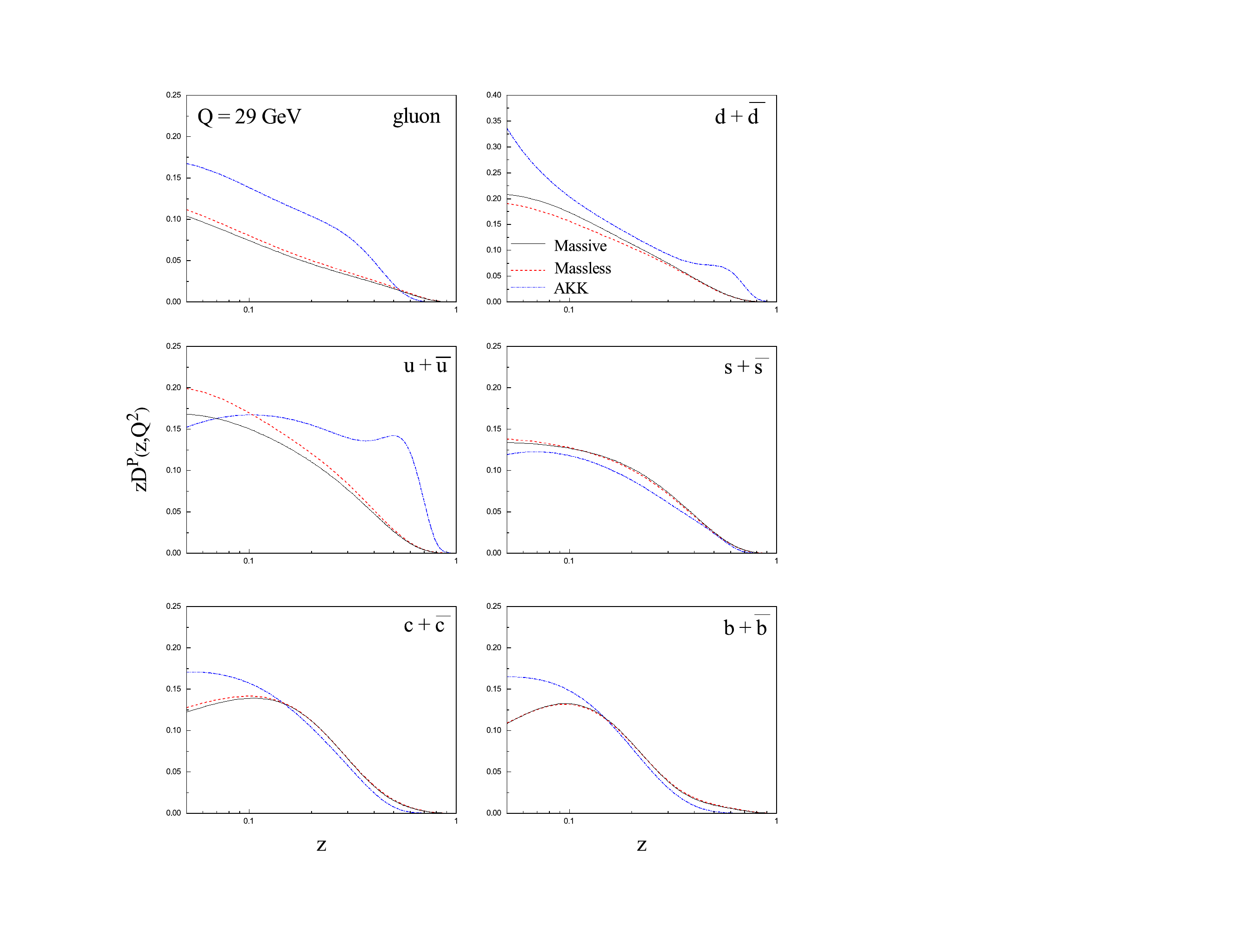}
		\caption{\label{PFFsQ29}%
			Comparison of the NLO proton FFs at $Q=29$ GeV for massive (solid lines) and massless (dotted lines) protons. The results are also compared
			with AKK \cite{Albino:2008fy} collaboration.}
	\end{center}
\end{figure}

\begin{figure}
	\begin{center}
		\includegraphics[width=1\linewidth,bb=80 50 500 550]{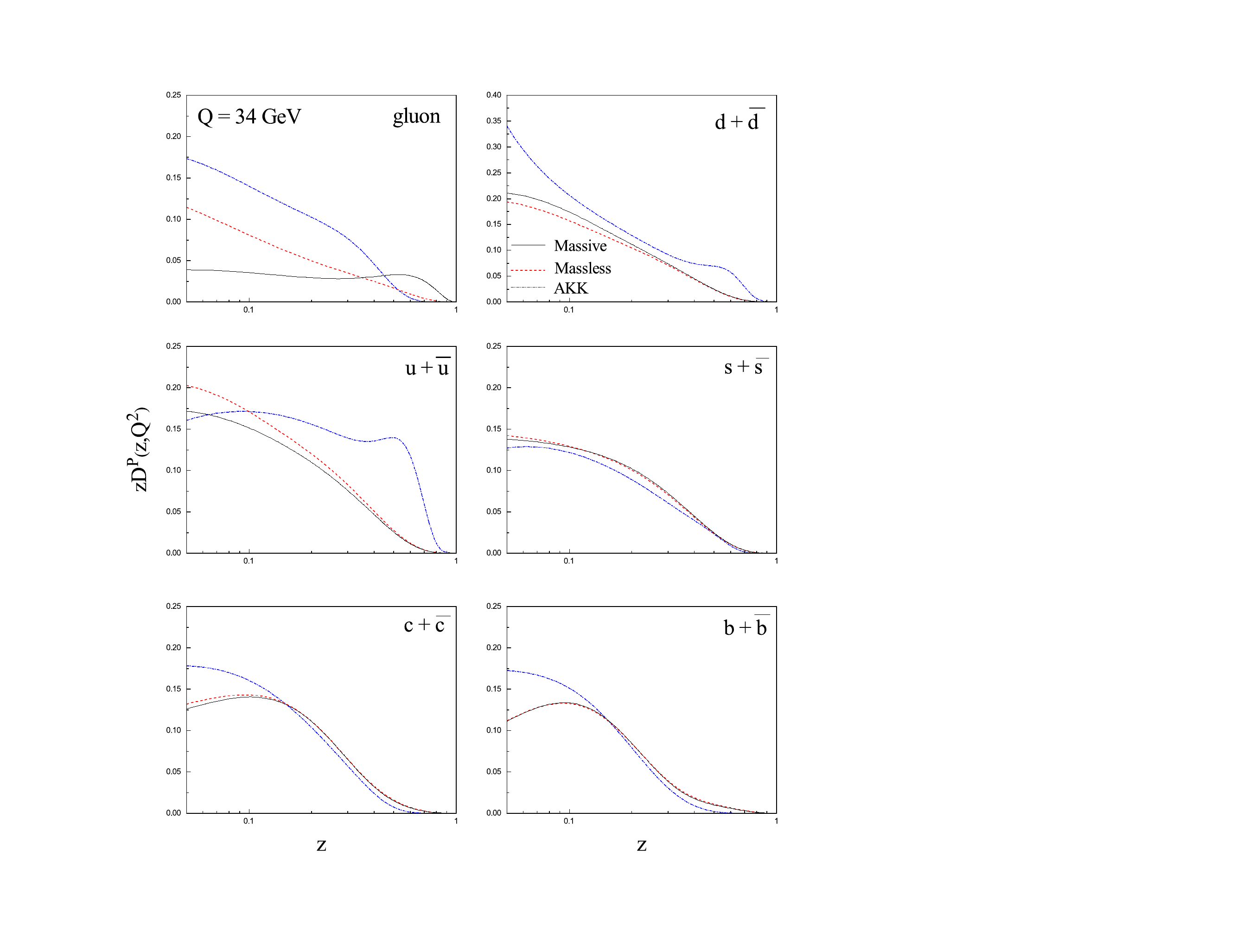}
		\caption{\label{PFFsQ34}%
			As in Fig.~\ref{PFFsQ29}, but at $Q=34$ GeV.}
	\end{center}
\end{figure}

\begin{figure}
	\begin{center}
		\includegraphics[width=1.2\linewidth,bb=120 35 780 535]{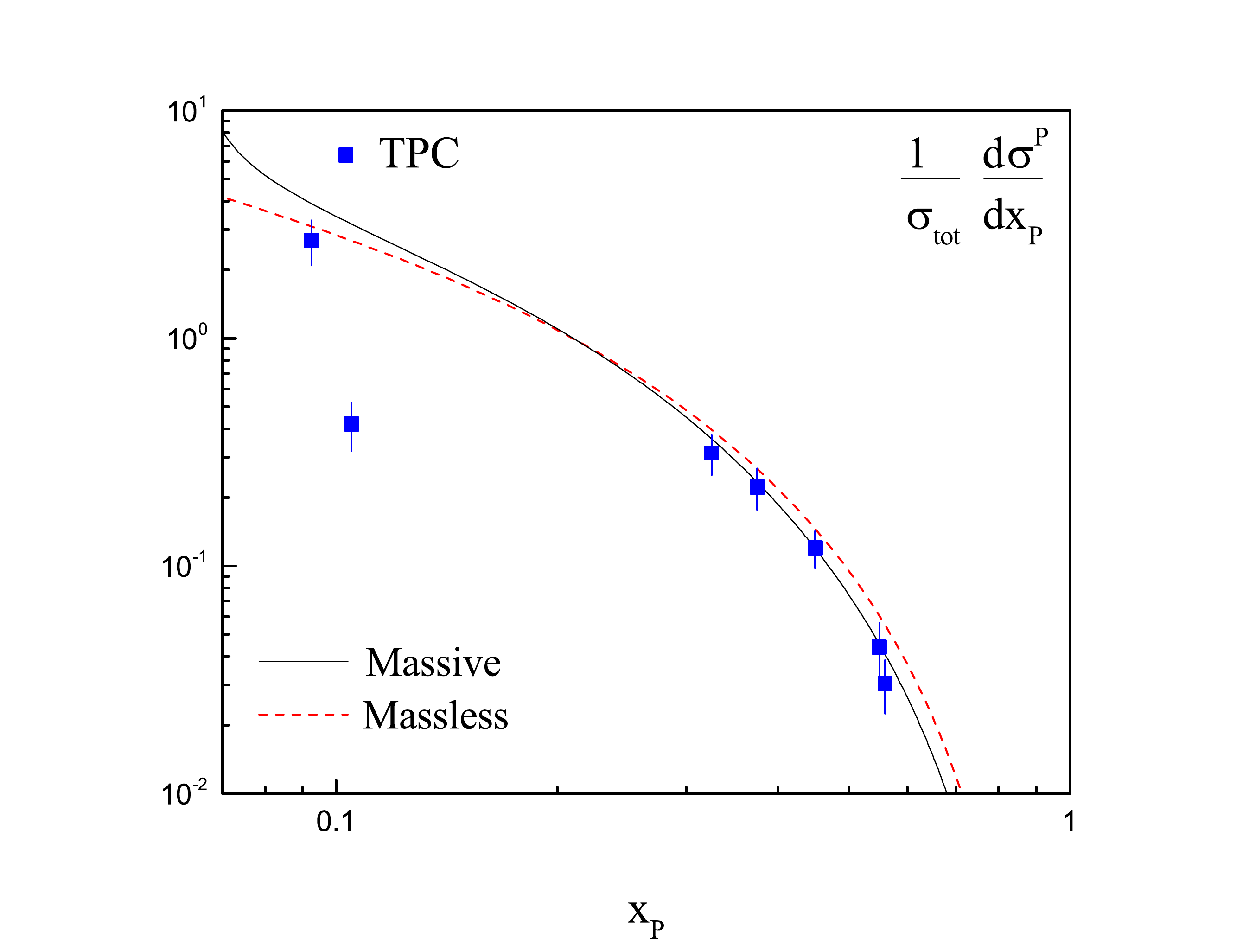}
		\caption{\label{SigQ29}%
			Comparison of our NLO results for the total cross section $1/\sigma_{tot}\times d\sigma/dx_p$ 
			with  data from {\it TPC} \cite{tpc29} at $Q=29$ GeV.
			The effect of proton mass is also considered in the solid curve.}
	\end{center}
\end{figure}

\begin{figure}
	\begin{center}
		\includegraphics[width=1.2\linewidth,bb=120 40 780 500]{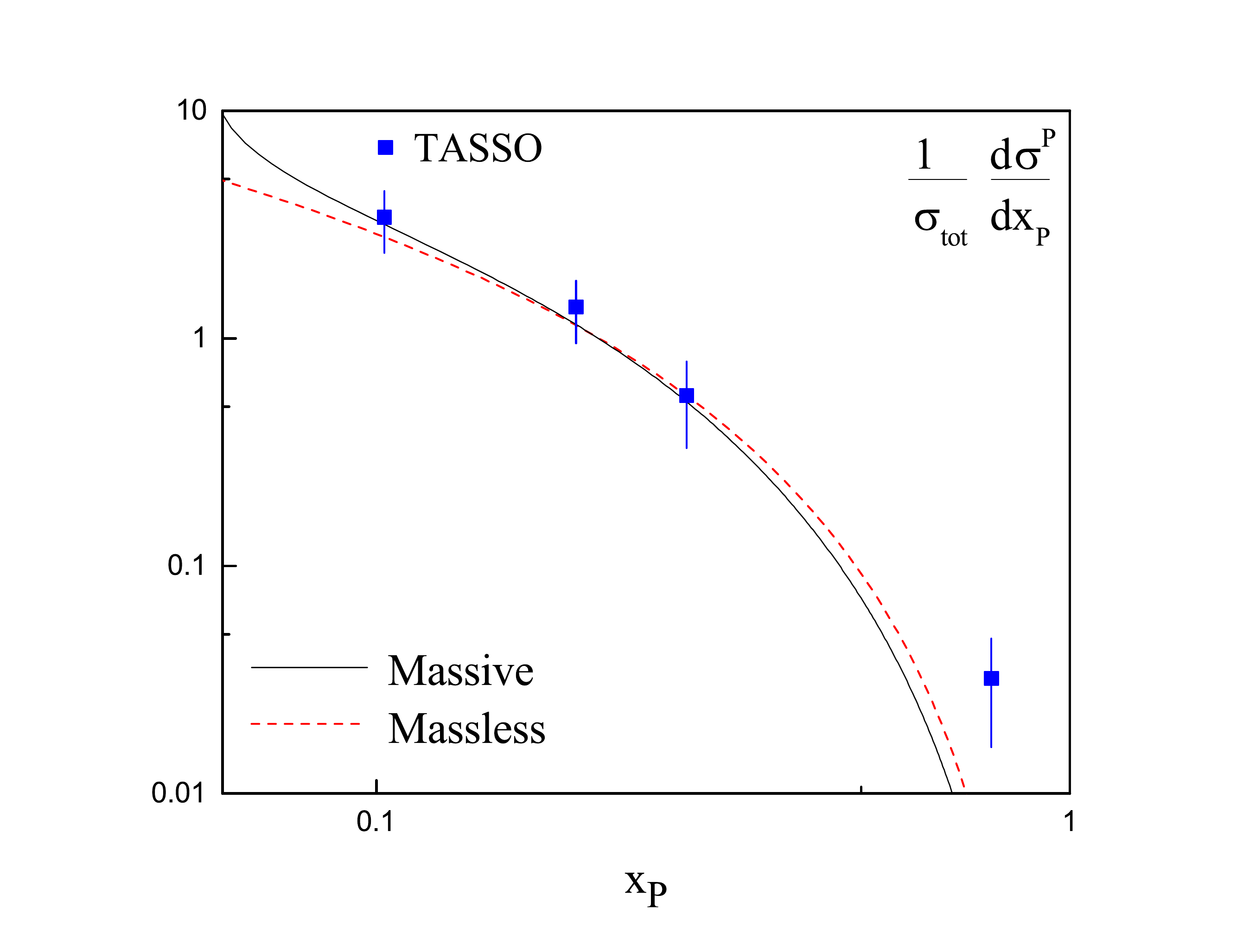}
		\caption{\label{SigQ34}%
			As in Fig.~\ref{SigQ29}, but comparison is done with data from {\it TASSO} \cite{tasso34_44} at $Q=34$ GeV.}
	\end{center}
\end{figure}

In this section, we show how to incorporate the hadron mass effects into 
inclusive hadron production in $e^+e^-$ reaction and into its relevant kinematic variables.
The same scheme can be applied for hadron-hadron reactions.
To explain our procedure, at first, we review the factorization theorem and its kinematic variables
defined. Consider the following process
\begin{eqnarray}\label{reac}
e^+e^-\rightarrow \gamma/Z(q) \rightarrow i(p_i)\bar{i}(p_{\bar{i}})\rightarrow H(p_H)+X,
\end{eqnarray}
where the four-momenta of particles are shown in the parenthesis, and $s=q^2$ is the squared of the center-of-mass (c.m.) energy.
 By ignoring the mass of produced hadron H
and outgoing partons ${\it i}$ and ${\it\bar i}$, in the c.m. frame the momenta take the following form 
\begin{eqnarray}\label{arash}
q=(\sqrt{s}, \vec{0}), \quad p_i=(E_i, \vec{0}, E_i), \quad p_H=(E_H, \vec{0}, E_H),\nonumber\\
\end{eqnarray}
 where we assumed that 
the parent partons, i.e. ${\it i}$ and ${\it\bar i}$, and produced hadron are emitted in the direction of the 3-axis.
Considering the definition of the cross section, one can write
\begin{eqnarray}
d\sigma(x_H, s)&=&\nonumber\\
&&\hspace{-2.2cm}\sum_{i=u, d, s, \cdots}{\int{dz d\hat{\sigma}_i(\mu_R^2, \mu_F^2)\Big|_{E_i=E_H/z} D_i^H(z, \mu_F^2)}},
\end{eqnarray}
where $d\hat{\sigma_i}$ are the Wilson coefficients in the parton level (the cross sections for the processes $e^+e^-\rightarrow i\bar{i}$)
and $D_i^H$ is the $i\rightarrow H$ FF. Here, $x_H=2p_H.q/q^2=2E_H/\sqrt{s}$ and $z=E_H/E_i$ are the scaling variables.
Since the measured observable is $d\sigma/dx_H$, by defining $x_i=2p_i.q/q^2=2E_i/\sqrt{s}$ one can write 
$d\hat\sigma_i=dx_H(dx_i/dx_H)d\hat\sigma_i/dx_i$, then
 \begin{eqnarray}
\frac{d\sigma}{dx_H}(x_H, s)=\sum_{i}{\int{dz \frac{d\hat{\sigma}_i}{dx_i}\frac{dx_i}{dx_H} D_i^H(\frac{x_H}{x_i}, \mu_F^2)}}.
\end{eqnarray}
Considering the definition $x_i=x_H/z$, one has $dx_i/dx_H=1/z$ and, therefore, one can write $dz/z=-dx_i/x_i$
so that the equation above is simplified  to the following form
\begin{eqnarray}\label{moadele}
\frac{d\sigma}{dx_H}(x_H, s)=\sum_{i}{\int_{x_H}^1{\frac{dx_i}{x_i} \frac{d\hat{\sigma}_i}{dx_i} D_i^H(\frac{x_H}{x_i}, \mu_F^2)}},
\end{eqnarray}
that leads to the relation (\ref{fac}). In the equation above, one has $\sqrt{\rho_H^2}\leq x_H\leq 1$ where $\rho_H=2m_H/\sqrt{s}$
is the cut-off and if one sets the hadron mass as $m_H=0$, then $0\leq x_H\leq 1$.\\
Note that, to establish the factorization theorem we ignored the hadron and parton 
masses when we defined the scaling variables $x_H$ and $x_i$.
To incorporate the hadron mass effects we use a specific choice of scaling variables.
For this purpose, it would be helpful to work with light-cone (L.C) coordinates, in which any four-vector $V$ is written in the
form of $V=(V^+,V^-,\vec{V}_T)$ with $V^\pm=(V^0\pm V^3)/\sqrt{2}$ and $\vec{V}_T=(V^1, V^2)$.\\
Considering the L.C coordinates, the four-momenta of particles in the $e^+e^-$ reaction (\ref{arash}) are expressed as
\begin{eqnarray}
q=(\frac{\sqrt{s}}{\sqrt{2}}, \frac{\sqrt{s}}{\sqrt{2}}, \vec{0}),  \quad p_H=(\sqrt{2} E_H, 0, \vec{0}),
\end{eqnarray}
and $ p_i=(\sqrt{2} E_i, 0, \vec{0})$.
Therefore, in absence of the hadron mass, the scaling variable $x_H=2E_H/\sqrt{s}$ is expressed as $x_H=p_H^+/q^+$ in the L.C coordinates.
From now on, in the presence of hadron mass, we define a new variable $\eta=p_H^+/q^+$ as a light-cone scaling variable which is identical to the $x_H$-variable in the absence 
of a hadron mass. The variable $\eta$ is now a more convenient scaling variable for studying hadron mass effects, because it is invariant with respect to
boosts along the direction of the hadron's spatial momentum ($Z$-axis). Taking a mass $m_H$ for the hadron, the momentum
of the hadron in the c.m. frame reads
\begin{eqnarray}
p_H=(p_H^+, p_H^-, \vec{p}_T)=(\eta q^+, p_H^-, \vec{p}_T)=(\eta\frac{\sqrt{s}}{\sqrt{2}}, p_H^-, \vec{0}).\nonumber\\
\end{eqnarray}
With respect to $p_H^2=m_H^2$ and considering the inner product in the L.C system ($V.V=2V^+V^--V_T^2$), the hadron momentum
is expressed as
\begin{eqnarray}\label{defen}
p_H=(\eta\frac{\sqrt{s}}{\sqrt{2}}, \frac{m_H^2}{\eta\sqrt{2s}}, \vec{0}).
\end{eqnarray}
As a generalization of the massless case, the cross section in the new coordinates is obtained by the 
replacements $x_H\rightarrow \eta(=p_H^+/q^+)$ and $x_i\rightarrow y(=p_i^+/q^+)$  in (\ref{moadele}), i.e.
\begin{eqnarray}
\frac{d\sigma}{d\eta}(\eta, s)=\sum_{i}{\int_{\eta}^1{\frac{dy}{y} \frac{d\hat{\sigma}}{dy} D_i^H(\frac{\eta}{y}, \mu_F)}}.
\end{eqnarray}
Since the experimental quantity is $d\sigma/dx_H$, it can be related to $d\sigma/d\eta$ via
\begin{eqnarray}
\frac{d\sigma}{dx_H}(x_H, s)=\frac{d\sigma}{d\eta}(\eta, s)\times\frac{d\eta}{dx_H}.
\end{eqnarray}
By comparing the hadron momentum in the L.C system, $p_H=((p_H^0+p_H^3)/\sqrt{2}, (p_H^0-p_H^3)/\sqrt{2}, \vec{0})$, 
with the equation (\ref{defen}), the equality  relation between two scaling variables results 
\begin{eqnarray}\label{ggh}
p_H^0=\frac{1}{2}(\eta\sqrt{s}+\frac{m_H^2}{\eta\sqrt{s}})\Longrightarrow x_H=\eta(1+\frac{m_H^2}{s\eta^2}).
\end{eqnarray}
Note that, these two variables are approximately equal when $m_h<<x_H\sqrt{s}$.
Considering equation (\ref{ggh}), one has
\begin{eqnarray}
\frac{d\eta}{dx_H}=\frac{1}{1-\frac{m_H^2}{s\eta^2(x_H)}}.
\end{eqnarray}
Finally, the differential cross section in the presence of  hadron mass reads
\begin{eqnarray}
\frac{d\sigma}{dx_H}(x_H, s)=\frac{1}{1-\frac{m_H^2}{s\eta^2(x_H)}}\frac{d\sigma}{d\eta}(\eta(x_H), s).
\end{eqnarray}
The above formula, established for the first time, would be 
a fundamental relation for the factorization theorem extended in the presence of hadron mass and it  would be more effective and applicable
when the hadron mass is considerable (especially,
larger than the proton mass).\\
Among all well-known collaborations, the only collaboration who studied the effects of proton mass into their calculations is AKK collaboration \cite{Albino:2008fy}.
The AKK have used the same coordinates to
include the effects of hadron mass in their analysis.
Due to the data used in their analysis, they have identified 
the scaling variable $x_m=2|\vec{p}|/\sqrt{s}$, where $|\vec{p}|$ stands for
 the three-momentum of the produced hadron, while in our analysis
 we use the energy scaling variable $x_H=2E_H/\sqrt{s}$. Therefore,   
the differential cross section (or the extended factorization formula), including the hadron mass effects computed in our work, is different with
the one ($d\sigma/dx_m$) presented in \cite{Albino:2008fy}. 
The effect of hadron mass is to reduce the size of the cross section $d\sigma/dx_m$ at small $x_m$, while this effect increases  the cross section $d\sigma/dx_H$ at small $x_H$. Besides, as was mentioned, the AKK has applied a simple parameterization form for the proton FFs by fixing some free parameters from  the beginning and they have also not determined the uncertainties of the proton FFs.\\
In Table.~\ref{tab:exmass},  we list all experimental data sets 
included in our global analysis, and the $\chi^2$ value per degree of freedom
pertaining to the NLO fit including the proton mass effects.
As is seen, the inclusion of the proton mass  
leads to a reduction of the value of $\chi^2 /d.o.f$, which is now $1.583$ in our global fit.
Our results for the fit parameters, in the presence of the proton mass, are listed in Table.~\ref{tab:mass}.
Using the Gaussian method,  their uncertainties are also shown.  
In Fig.~\ref{plotmodel3}, using the fit parameters presented in Table.~\ref{tab:mass}, the massive proton FFs (solid line)
are shown at the scales $Q^2=2$ GeV$^2$, $4$ GeV$^2$ and
$25$ GeV$^2$ for the gluon and light quarks, the  {\it c}- and {\it b}-quarks, respectively. Considering errors presented in Table.~\ref{tab:mass}, the uncertainties of the massive proton FFs are also shown.
In Figs.~\ref{PFFsQ29} and \ref{PFFsQ34}, the effect of proton mass on the parton FFs are shown 
at the scales $Q=29$ GeV and $Q=34$ GeV, respectively. The massless (dotted line) and massive (solid line) FFs are also 
compared with the results obtained by AKK collaboration \cite{Albino:2008fy}, who  used the proton mass effects too.
As it is seen, the mass of the proton affects the FFs
of the gluon and light quarks $d$ and $u$.
It is also seen that our results for the gluon,  {\it d-} and {\it u-}quark FFs deviate from the AKK extractions of FFs; however the data sets and the extended factorization theorem in their analysis are different.
In Fig.~\ref{SigQ29}, our results for $1/\sigma_{tot}\times d\sigma^p/dx_p$ in the presence of  massless and massive protons are compared with the data from the {\it TPC} collaboration at $Q=29$ GeV.
In Fig.~\ref{SigQ34}, the same comparison is done  with the data from the {\it TASSO} collaboration at $Q=34$ GeV. 
Note that the effect of the proton mass is that it reduces the size of the differential cross section at large $x_p$. This is needed to improve the fit when one works with data from {\it TPC}. As is shown, the effect of the proton mass is that it increases the size of the cross section at small $x_p$ so that improves the fit when we work with the data from the {\it TASSO} collaboration, see Fig.~\ref{SigQ34}.
\begin{table}[h]
	\caption{\label{tab:exmass}The individual $\chi^2$ values  for
		each collaboration and the total $\chi^2$ fit for proton when including the proton mass.}
	\begin{ruledtabular}
		\tabcolsep=0.05cm \footnotesize
		\begin{tabular}{lccccc}
			collaboration& data & $\sqrt{s}$  GeV&  data   &  $\chi^2$(NLO)\\
			& properties &          &  points & normalization in fit &   
			\\\hline
			TPC \cite{tpc29}  &   untagged   & 29      & 8  & 20.648  \\
			TASSO \cite{tasso34_44}  & untagged \  &  34 & 4 & 2.118\\
			ALEPH \cite{aleph91}    & untagged  & 91.2 & 18 &  18.408 \\
			SLD \cite{sld91}  & untagged\  & 91.28 & 28  &83.724\\
			& $uds$ tagged         &  91.28 & 29 & 42.743 \\
			& $c$ tagged           &  91.28 & 29 & 38.603\\
			& $b$ tagged           &  91.28 & 28 & 43.927\\
			DELPHI \cite{delphi91,delphi91-2}  & untagged & 91.2  & 17 & 4.390 \\
			& $uds$ tagged         &  91.2 & 17 & 4.994 \\
			& $b$ tagged           &  91.2 & 17 & 14.241 \\
			{\bf TOTAL:} & & &195 &273.796\\
			($\chi^{2}$/ d.o.f ) & & & &1.583\\
		\end{tabular}
	\end{ruledtabular}
\end{table}
\begin{table}[h]
	\caption{\label{tab:mass} 
		Values of the fit parameters for the  proton FFs at NLO, by including proton mass effects in the fit. We set
		the proton mass as $m_p=938.272$ MeV.}
	\centering
	\tabcolsep=0.05cm \footnotesize
	\hspace{1cm}
	\begin{tabular}{cccccc}
		\hline
		\hline
		flavor $i$ &$N_i$ & $\alpha_i$ & $\beta_i$ &$\gamma_i$\\
		\hline
		$u$ &$2.016\pm 2.189$&$-0.695\pm 0.165$&$16.080\pm 2.329$&$2.314\pm 0.152$ 
		\\
		$d$ &$4.676\pm 1.879$&$-0.695\pm 0.165$&$16.080\pm 2.329$&$2.314\pm 0.152$  
		\\
		$ \overline{u}$&$19.005\pm 2.002$&$-0.601\pm 0.070$&$3.665\pm 1.106$&$0.090\pm 0.017$  
		\\
		$ \overline{d}$&$17.984\pm 3.707$&$-0.601\pm 0.070$&$3.665\pm 1.106$&$0.090\pm 0.017$
		\\
		$s, \overline{s}$&$8.878\pm 1.809$&$-0.601\pm 0.070$&$3.665\pm 1.106$&$0.090\pm 0.017$
		\\
		$c, \overline{c}$&$6.908\pm 2.042$&$0.867\pm 0.137$&$7.967\pm 2.069$&$...$
		\\
		$b, \overline{b}$&$7.560\pm 2.176$&$0.569\pm 0.169$&$11.569\pm 1.284$&$...$ 
		\\
		$g$&$3.515\pm 1.154$&$4.388\pm 0.433$&$1.739\pm 0.117$&$...$   
		\\
		\hline
		\hline
	\end{tabular}
\end{table}
\section{Energy spectrum of the inclusive proton in top quark decays}
\label{sec4}
\begin{figure}
	\begin{center}
		\includegraphics[width=1.1\linewidth,bb=20 190 600 635]{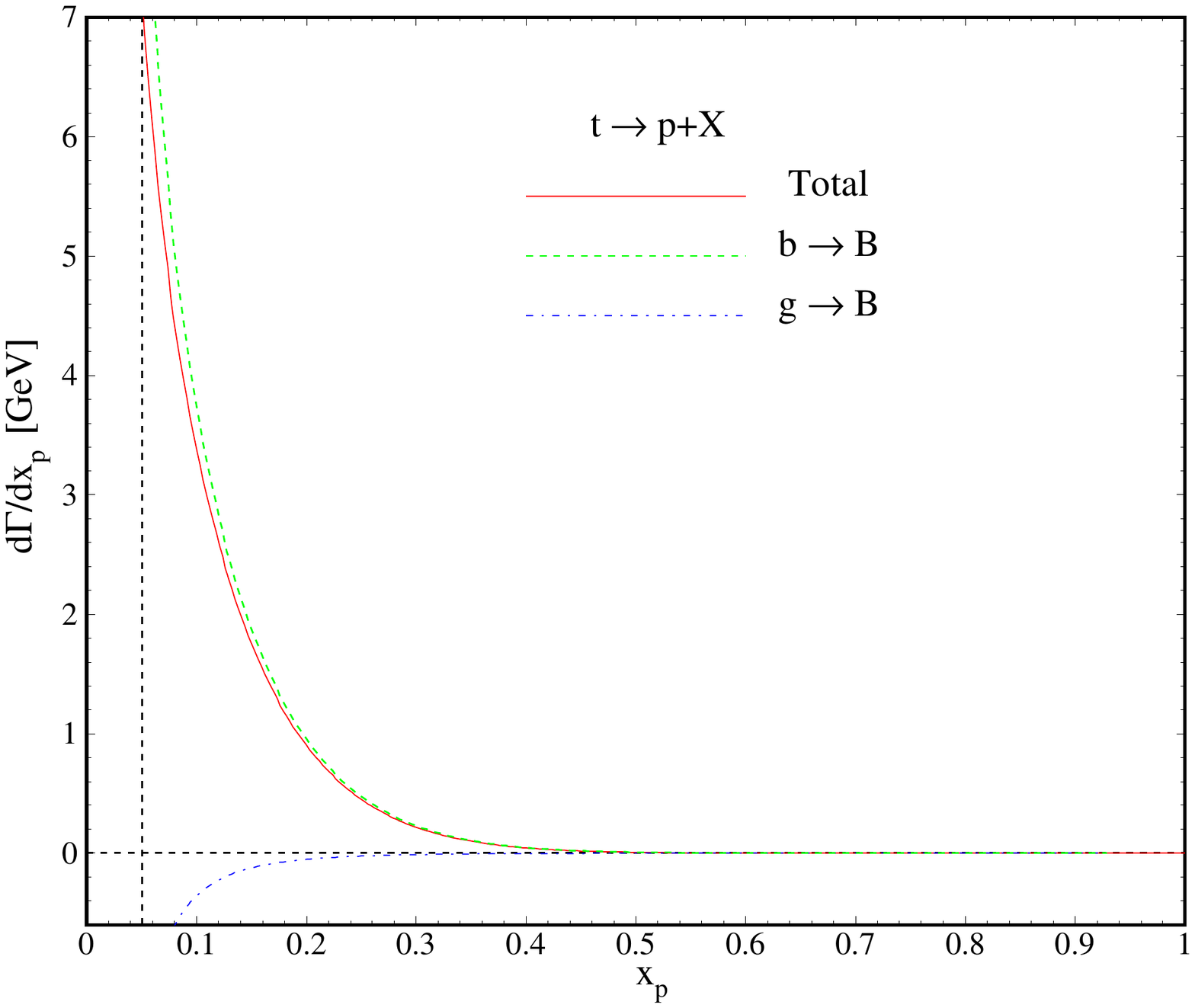}
		\caption{\label{fig10}%
			$d\Gamma(t\rightarrow p+X)/dx_{p}$ as a function of $x_p$ (solid line)
			at $\mu_F=m_t$. The NLO result is broken up into the
			contributions due to $b\rightarrow p$ (dashed line) and $g\rightarrow p$ (dot-dashed line) fragmentations.}
	\end{center}
\end{figure}

\begin{figure}
	\begin{center}
		\includegraphics[width=1.1\linewidth,bb=20 190 600 635]{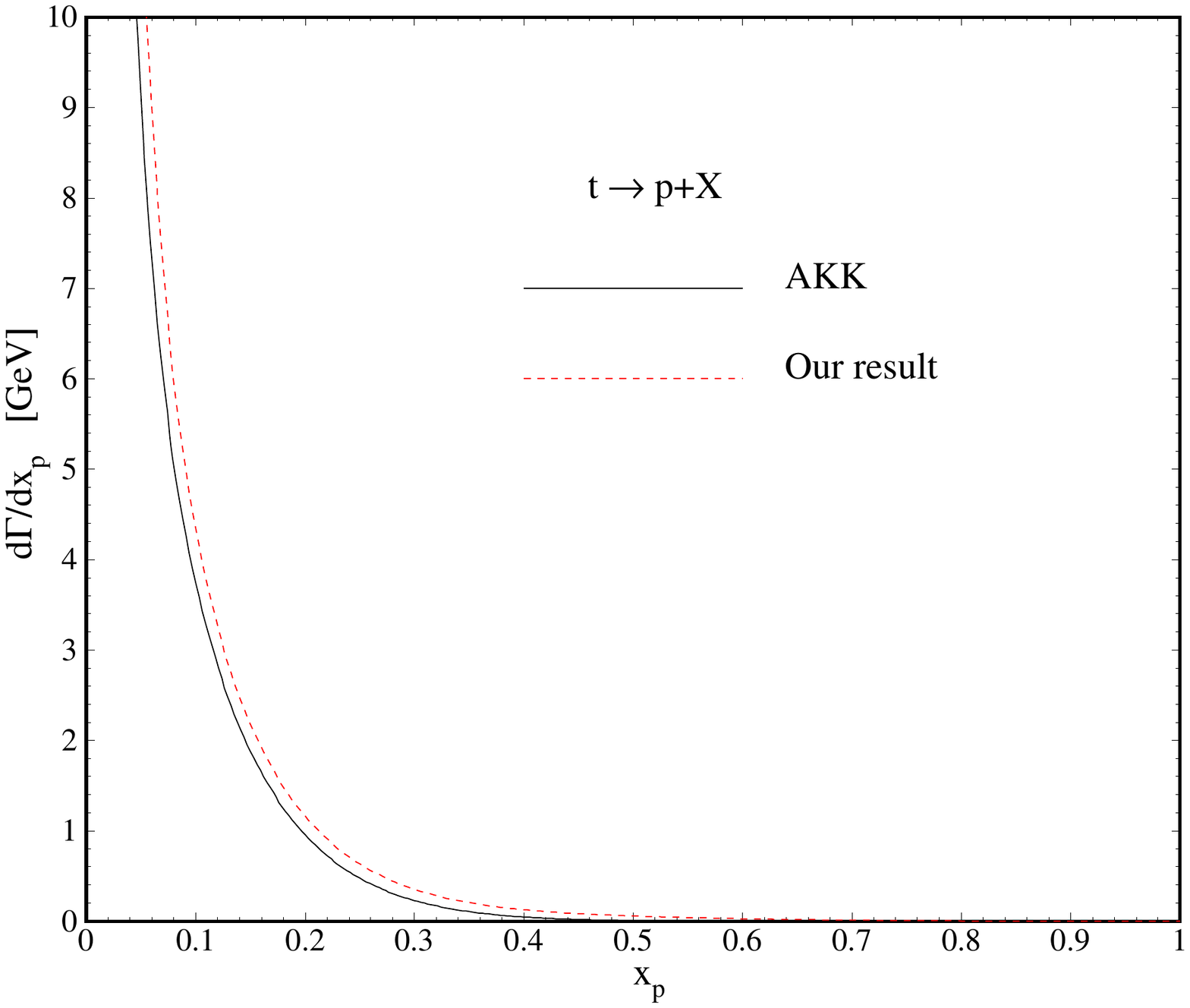}
		\caption{\label{fig20}%
			Result for the partial decay rate $d\Gamma/dx_p$
			is shown using the extracted $(b, g)\rightarrow p$ FFs in the presence of the proton mass.
			We also show the same distribution using the FFs presented by AKK \cite{Albino:2008fy}.}
	\end{center}
\end{figure}

In this section, we apply the massive proton FFs  to make phenomenological predictions for the energy spectrum of protons produced in  unpolarized top decays
\begin{eqnarray}\label{pros}
t\rightarrow b+W^+ (g)\rightarrow p+X,
\end{eqnarray}
where $X$ stands for the unobserved final state.  Both the $b$-quark and the gluon may hadronize into 
the produced protons,  while the gluon contributes to the real radiation at NLO.\\
Generally, to obtain the energy distribution of the produced hadron $H$, we employ the factorization theorem  (\ref{moadele}), which is now expressed as 
\begin{eqnarray}\label{eq:master}
\frac{d\Gamma}{dx_H}=\sum_{i=b,g}\int_{x_i^{min}}^{x_i^{max}}
\frac{dx_i}{x_i}\,\frac{d\hat\Gamma}{dx_i}(\mu_R,\mu_F)
D_i^H\left(\frac{x_H}{x_i},\mu_F\right),\nonumber\\
\end{eqnarray}
where, as in \cite{Corcella:2001hz}, we defined the scaled-energy fraction of the hadron as $x_H=2E_{H}/(m_t^{2}-m_W^{2})$   and
$d\hat\Gamma/dx_i $ are the parton-level differential decay rates of the process
$t\to i+W^+ (i=b,g)$. The NLO analytical expressions for the parton-level differential decay widths
$d\hat\Gamma/dx_i$  are presented in Refs.~\cite{Kniehl:2012mn,Corcella:2001hz}.
In \cite{Nejad:2016epx}, we studied the effects of parton and  hadron masses on the hadron energy spectrum produced through top decays.\\
In (\ref{eq:master}), the factorization and the renormalization scales are set to $\mu_R=\mu_F=m_t=172.9$ GeV and we also set
$m_b=4.78$ GeV, $\Lambda^{NLO}=231$ MeV and $m_W=80.39$ GeV.\\
In Fig.~\ref{fig10}, to show our prediction for the energy spectrum of produced protons  we study  the size of the NLO
corrections, by showing the total result (solid line), and comparing the relative
importance of the $b\rightarrow p$ (dashed line) and $g\rightarrow p$ (dot-dashed line) fragmentation channels. As is seen, the gluon contribution (dot-dashed line) is negative and appreciable only in the low $x_p$ region.
For higher values of $x_p(x_p>0.3)$ the NLO result is
practically exhausted by the $b\rightarrow p$ contribution, as expected \cite{Corcella:2001hz}.
Note that the contribution of the gluon FF cannot be discriminated. It is
calculated to see where it contributes to $d\Gamma/dx_p$. 
So, this part of the plot is of more theoretical relevance. In the
scaled-energy of hadrons, as an experimental quantity, all
contributions including the b-quark and gluon
contribute.
In Fig.~\ref{fig20}, our prediction for  $d\Gamma(t\rightarrow p+X)/dx_p$ is compared with the result
obtained by using the FFs extracted  by the AKK collaboration \cite{Albino:2008fy}. Considering Figs.~\ref{PFFsQ29} and \ref{PFFsQ34}, it is seen that our results for the b-quark fragmentation function are in good consistency with the results extracted by the AKK for the values higher than $z\geq 0.1$, then one expects to have the same results for the $d\Gamma/dx_p$ in both analysis.


\section{Conclusion}
\label{sec5}
We determined new nonperturbative fragmentation functions for the proton  in the parton
model of QCD through a global analysis on SIA  data at  the LO and NLO approximations.
Our new parametrization of the proton FFs covers a wide kinematic range of $z$, in the presence of the extra term $(1-\delta_i e^{-\gamma_i z})$
which controls the medium-$z$ region and improves the accuracy of the global fit.
Fig.~\ref{xsec3} represents the comparison of our model with SIA experimental data
and shows that our model is successful. We determined the FFs of gluon and light quarks at the initial scale $\mu_0^2=1$ GeV$^2$
and the FFs of heavy quarks at $\mu_0^2=m_c^2$ and $\mu_0^2=m_b^2$. Their evaluation was performed
by using the DGLAP equations.
Our analysis was based on the ZM-VFN scheme, where all partons
are treated as massless particles and 
the nonzero values of the c- and b-quark masses only enter through the initial conditions of the FFs.
Comparing to  other collaborations  we also considered the effects of the proton mass on the FFs
and showed that this effect improves the accuracy of the global fit, specifically for the data from SLAC (TPC collaborations) and DESY (TASSO collaboration).
The proton mass affects the light and gluon FFs while for the heavy quark FFs this effect is less important.
The advent of precise data from LHC offers us the opportunity to further constrain the proton FFs
and to test their scaling violations. This situation motivates the incorporation of proton mass effects into the formalism, which are
then likely to be no longer negligible.
The mass of the proton also sets a bound on the scaling variable $x_p\geq2m_p/\sqrt{s}$.\\
Finally, as an application, we used the computed FFs to study the scaled-energy  distribution of  protons
in unpolarized top quark decays. At LHC, the scaled-energy distribution of hadrons in top  decays  enables us to deepen our
knowledge of the hadronization process. The universality and scaling violations of the proton FFs will be able to be tested 
at the LHC by comparing our NLO predictions with future measurements of $d\Gamma/dx_p$.

Besides NLO corrections, some other corrections can improve our analysis.
In future work, we plan to extend the method developed in \cite{Cacciari:2001cw} in the framework of $e^+e^-$ annihilation to resum large logarithmic terms, due to soft-gluon radiation, to all perturbative orders in the QCD coupling
$\alpha_s$.

\begin{acknowledgments}
	We are grateful to Gennaro Corcella for his careful reading of the manuscript
	and also for constructive discussions and comments.
	We would also like to thank  Mathias  Butenschon for reading and editing the English
	manuscript and Bernd A.~Kniehl and the DESY
	theoretical division for their hospitality,
	where a portion of this work was performed.
\end{acknowledgments}


\end{document}